\documentclass[pdflatex,sn-mathphys-num]{sn-jnl}

\usepackage{graphicx}%
\usepackage{multirow}%
\usepackage{amsmath,amssymb,amsfonts}%
\usepackage{amsthm}%
\usepackage{mathrsfs}%
\usepackage[title]{appendix}%
\usepackage{xcolor}%
\usepackage{textcomp}%
\usepackage{manyfoot}%
\usepackage{booktabs}%
\usepackage{algorithm}%
\usepackage{algorithmicx}%
\usepackage{algpseudocode}%
\usepackage{listings}%

\theoremstyle{thmstyleone}%
\theoremstyle{thmstyletwo}%

\theoremstyle{thmstylethree}%

\begin{document}

\title[Article Title]{Distributed Acoustic Sensing for Urban Monitoring: Coverage Thresholds and Percolation}


\author[1]{\fnm{Khen} \sur{Cohen}} 
\author[2]{\fnm{Ariel} \sur{Lellouch}}
\affil[1]{\orgdiv{The School of Physics and Astronomy}, \orgname{Tel-Aviv University}, \orgaddress{\street{Haim Levanon}, \city{Tel Aviv}, \postcode{69978}, \country{Israel}}}
\affil[2]{\orgdiv{Department of Geophysics}, \orgname{Tel-Aviv University}, \orgaddress{\street{Haim Levanon}, \city{Tel Aviv}, \postcode{69978}, \country{Israel}}}


\abstract{
Distributed Acoustic Sensing (DAS) enables the repurposing of existing fiber-optic networks as ultra-dense, long-range seismic arrays for urban monitoring. However, constraints imposed by real-world fiber infrastructure topology and components limit its use for city-scale applications. Recent technological developments have paved the way for short-range, on-chip DAS. Assuming their availability, and based on a Graph Theory framework, we show that monitoring applications fall along a coverage spectrum with two critical thresholds that define three distinct regimes. Low coverage ($<10 \%$) can, with optimal design, resolve earthquake early warning, groundwater monitoring, geological mapping, and urban activity tracking. A percolation transition occurs at $\approx 51.6\%$ coverage, beyond which the city effectively becomes fully covered and statistical traffic monitoring is possible. Only for effectively complete coverage, infrastructure monitoring, individual vehicle tracking, and pedestrian movement analysis become possible. Thus, privacy-related risks remain very low. We show and exemplify how, for metropolises around the world, an optimal sensing network can be designed for earthquake early warning, traffic monitoring, and urban activity tracking. This framework provides a near-future roadmap for deploying urban DAS networks as a backbone of smart city sensing.}

\keywords{Distributed Acoustic Sensing (DAS), Smart Cities, Percolation Theory, Urban Mobility, Seismic Resilience, Privacy-Preserving Sensing}



\maketitle
\section{Introduction}
\label{sec:intro}
Urban science utilizes quantitative, data-driven approaches to support rapid urbanization by informing policy makers, promoting sustainability, and improving citizens' quality-of-life and safety using a cross-disciplinary view \cite{acuto2018building,Bettencourt2021}. Traditionally, measurements are obtained primarily from dedicated point sensors that target different components of urban life, such as mobility, water quality, and energy usage \cite{ramirez2021sensors}. However, deploying such sensors on a city scale incurs significant deployment and maintenance efforts and is thus inherently limited in the achievable spatial resolution. Furthermore, privacy concerns associated with video surveillance increasingly limit the deployment of visual monitoring systems \cite{guerrero2018sensor, kitchin2016ethics}. 

Recently, seismic signals have been shown to contain valuable information \cite{diaz2017urban} that can be used to cost-effectively monitor urban environments while maintaining privacy. Among others, they can be used to follow traffic patterns \cite{Lindsey2020COVID, cohen2025fiber}, monitor infrastructure health \cite{zhu2022infrastructure}, detect shallow subsurface changes \cite{yuan2024DAS,Fang2020Urban} and groundwater variations \cite{Li2025groundwater}, and track levels of urban activity \cite{Lecocq2020COVID, liu2025urbanreview}. Among seismic sensors, Distributed Fiber Optic Sensing (DFOS) and in particular Distributed Acoustic Sensing (DAS) have been shown to be highly effective in urban environments \cite{liu2025urbanreview}. DAS interrogators utilize Rayleigh backscattered light to infer the strain or strain-rate along the fiber \cite{hartog2017introduction}. It offers quasi-continuous spatial measurements for distances of tens of kilometers or more, allows leveraging existing telecommunication infrastructure without additional deployments, and is sensitive enough to detect signals as faint as human footsteps and rats in optical conduits \cite{wang2025rats,luckie2025footsteps}. However, the relationship between urban fiber-network topology and optimal DAS architecture has remained largely unexplored, and used fiber routes are generally driven by access and logistical considerations.

For the past decade, developments in DAS have improved signal-to-noise ratio, increased sensing range, and lowered interrogator costs \cite{chen2019range, fernandez2022seismic, he2021optical,waagaard2021range171,fan2023range300,markom2025systematic}. Nonetheless, to leverage these improvements, DAS requires long uninterrupted fiber-optic cables. The topology of existing urban telecommunications networks typically has the opposite structure, with frequent filtering, switching, and routing nodes that disturb the backscattered light required for DAS \cite{simmons2014network}. Therefore, it is practically challenging to create a continuous fiber route in an urban environment, and doing so requires many optical modifications that operators generally wish to avoid such as fiber splicing or optical cabinet alterations. In addition, the eventual purpose of fiber-optic networks is to reach individual buildings or homes (Fiber To The Home - FTTH). This lends their design to a tree-like structure at the access network, or last-mile, level \cite{ramaswami2009optical}. As a direct consequence, creating continuous fiber routes for urban sensing becomes impractical as it requires going back and forth along the same route after optical modifications at multiple end points. Therefore, long-range DAS interrogators are, in fact, suboptimal for urban sensing. 

During the last years, however, significant progress has been made in the realm of miniaturized DAS sensing \cite{jin2024silicon,cheng2023chip,wu2025compact}. This technological direction offers low-cost, on-chip DAS measurements. Whereas their expected signal quality and range is lower than existing interrogators, they have the benefit of being potentially deployed along selected segments of the urban fiber-optic network without requiring optical modifications to create a long contiguous optical route. Here, we show that the topology of urban fiber-optic networks fundamentally favors distributed short-range DAS architectures. Using a Graph Theory framework and percolation theory, we analyze the potential of a distributed network of DAS (DDAS) for a variety of applications such as earthquake early warning, traffic monitoring, and urban activity tracking. We also show and exemplify how the DDAS network can be optimized for different tasks in major cities across the world. We show that even under low coverage conditions ($\approx 10 \%$ or less), optimal design of DDAS can lead to substantial benefits. 

\begin{figure}[htbp]
    \centering
    \includegraphics[trim={0cm 3.5cm 0cm 3cm}, clip, width=1.0\textwidth]{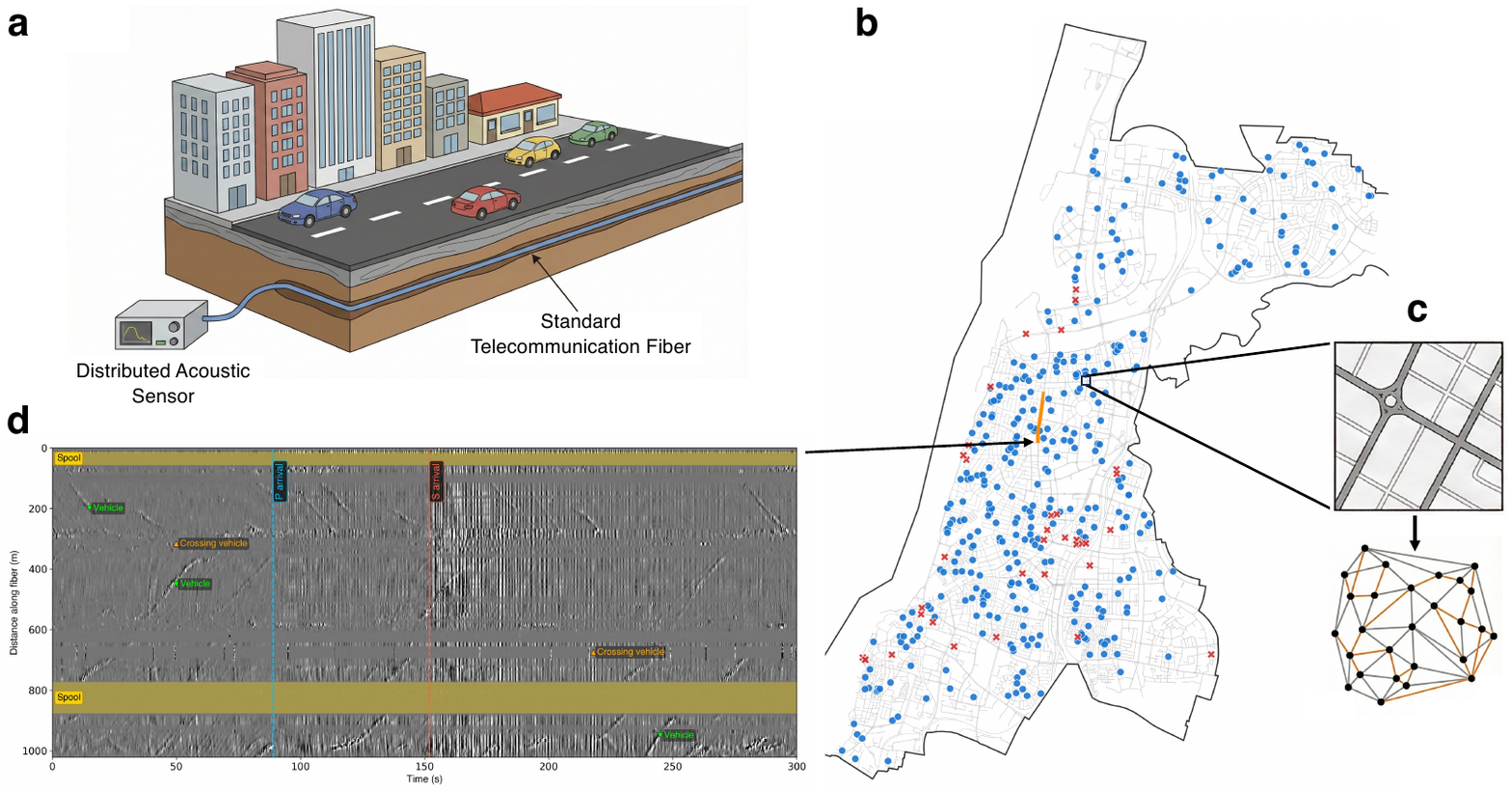}
    \caption{\textbf{Distributed Acoustic Sensor at city scale}. \textbf{a} Schematic of the urban sensing principle. A standard telecommunication fiber (active or passive) is interrogated by a laser unit, converting the cable into a dense array of strain sensors. The system detects geophysical signals from seismic events and can monitor both traffic and seismic activity. \textbf{b} Optical fiber coverage in the Tel Aviv municipality (see Supplementary Materials for the full analysis) supports the conclusion that the fiber infrastructure covers most of the city streets. \textbf{c} The urban street map (top) is modeled as a planar graph $\boldsymbol{G}(V,E)$ (bottom), where road junctions form nodes $V$ and street segments form edges $E$. Fiber-optic routes are represented as a subset $e \subseteq E$ (orange edges). \textbf{d} DAS recording in Tel Aviv along Ibn Gvirol St. (orange line), showing simultaneous earthquake detection and traffic monitoring.}
    \label{fig:concept}
\end{figure}

\section{Results}
\label{sec:results}

\begin{figure}[t]
    \centering
    \includegraphics[trim={2cm 2.5cm 2cm 2cm}, clip, width=0.9\textwidth]{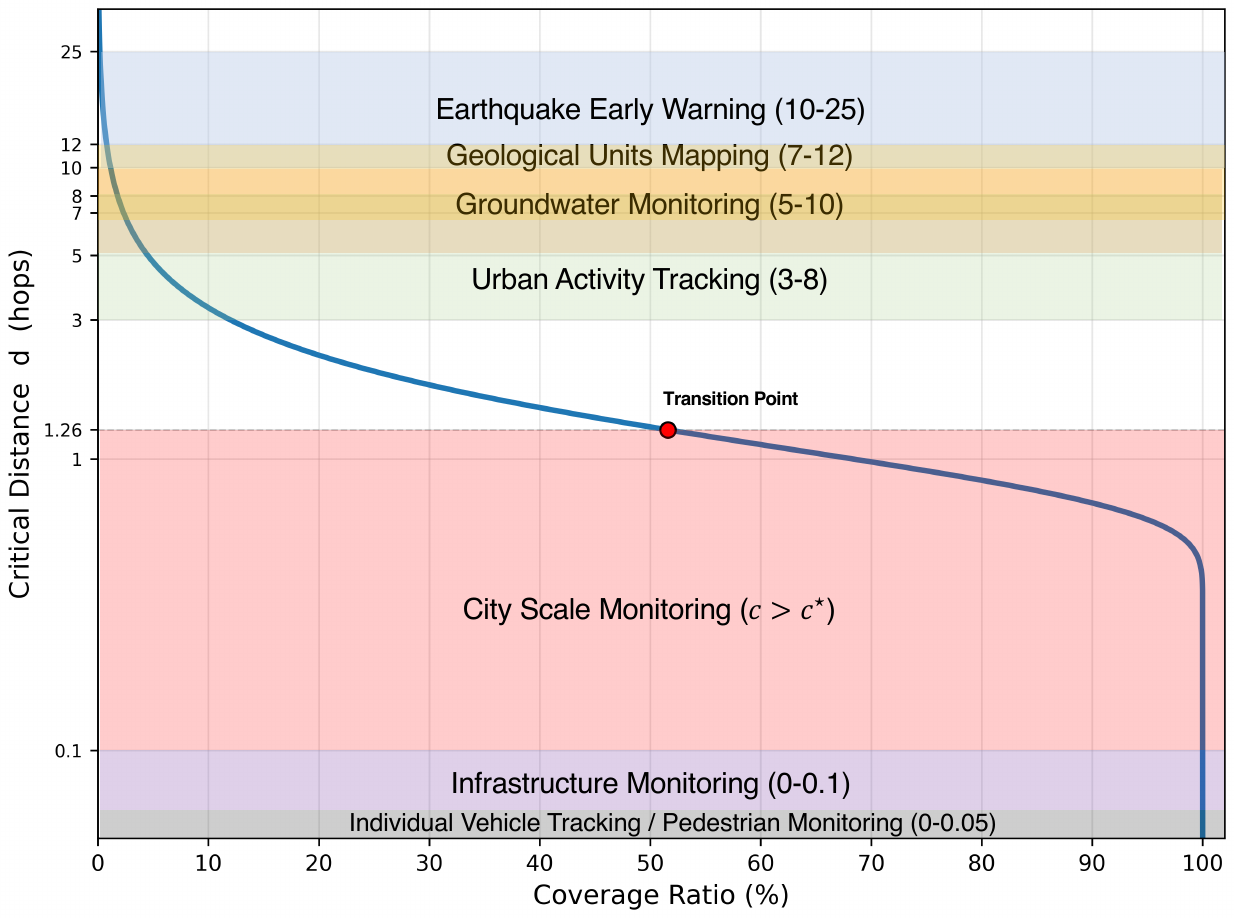}
    \caption{\textbf{Critical distance as a function of the coverage ratio}, according to eq. \ref{eq:dc}, with success probability of $P_{\text{success}}=90\%$. The city coverage phase transition point (red circle) at $\boldsymbol{c}^\star \approx 51.6\%$. This curve establishes the connection between the coverage ratio and the distance length, for a typical urban area. The units of $d$ are in topological distance, and it is derived from the km-units distance as shown in the Methods.}
    \label{fig:scalingPerc}
\end{figure}

\subsection{Mapping and Representing the Topology of Urban Fiber Optic Networks}
Generally, the exact structure and layout of optical networks is not shared by telecommunication operators. Large-scale studies \cite{adib2025network,fcc2026acess} estimate coverage of between $40-75 \%$ of all homes in major countries (UK, Germany, USA, Canada). These values are typically higher in dense urban areas and lower in rural environments. It is possible, nonetheless, to verify whether a certain address is connected to fiber using customer websites. Thus, to exemplify typical coverage in dense urban environments, we map fiber access in the Tel Aviv-Jaffa metropolitan area using queries on telecommunication operators' websites. Figure \ref{fig:concept} illustrates mapped fiber access areas. We find that the majority of streets are covered by fiber on at least one of their side. Uncovered streets fall into two categories: narrow streets that are non-residential or sparsely populated, and large roads with no immediate neighboring population. We emphasize that this map is the worst case scenario as it represents private customer access only. It is possible, for instance, that the fiber infrastructure runs along major roads with no adjacent residential areas, and therefore service is not offered despite existing fiber. This analysis shows that we can reasonably assume that every street in a city is covered by fiber. 

Therefore, we can use random Planar Graphs to represent the topology of fiber-optic networks directly from road maps (see Methods for justification). Each street segment is an edge and each junction is a node. We also assume that advances in short-range DAS will allow us to choose any subset of edges from the graph to record data over. Under this framework, we analyze and optimize the recording network for different applications that DFOS has been shown valuable for. For generalization, we present statistics of urban areas worldwide (See Supplementary Materials) indicating that a typical city has from several thousand to a million streets, with an average of approximately $N = 17{,}000$ junctions, a value we use for the percolation analysis. 

\subsection{Critical Distance and Percolation Analysis}
We analyze the planar graph representing every city with respect to the order parameter $c$, the \textit{coverage ratio}, which is defined as the ratio between the number of streets (edges) with DAS sensing to the total number of streets. Assuming mean field approximation with some constant density, the probability that we can find another interrogated fiber in an area with surface $\propto d^2$ is referred as the \textit{critical distance}, and is given by $d=\sqrt{\tfrac{1}{2} log_{1-\boldsymbol{c}}(1-P_{\text{success}})}$. We assume that $P_{\text{success}}$, which represents the average success probability of each application, is set to 90\%,  and show the required coverage ratio to achieve different values of $d$. Scaling between $d$, a dimensionless number, and physical distance depends on average ratio between the number of streets and the city total area (which was evaluated based on the dataset we used). We use this value to estimate the required values for different applications (See Methods) and summarize the results in Figure \ref{fig:scalingPerc}. 

We observe three different regimes. The first is a low-coverage scenario ($\boldsymbol{c} \leq 0.1$), which is sufficient for large-scale processes that do not require spatial sensing continuity (earthquake early warning, groundwater monitoring, geological mapping, urban activity monitoring). The second is street-level coverage ($\boldsymbol{c} \rightarrow 1$), which is relevant to applications for which the DAS fiber needs to be very close to the target. In between, we have a transition regime. For $\boldsymbol{c}^\star \approx 0.516$, the percolation threshold is reached, and the DAS network becomes a Giant Connected Component (GCC), as also verified numerically in the Supplementary Materials. Under these conditions, we can claim that the city-scale coverage is attained by the DAS, and we can apply any type of city-wide monitoring, such as statistical traffic estimation, without the need to track individual vehicles nor cover every road in the city. 

\begin{figure}
    \centering
    
    \includegraphics[trim={0cm 3.5cm 0cm 2.5cm}, clip, width=1.0\textwidth]{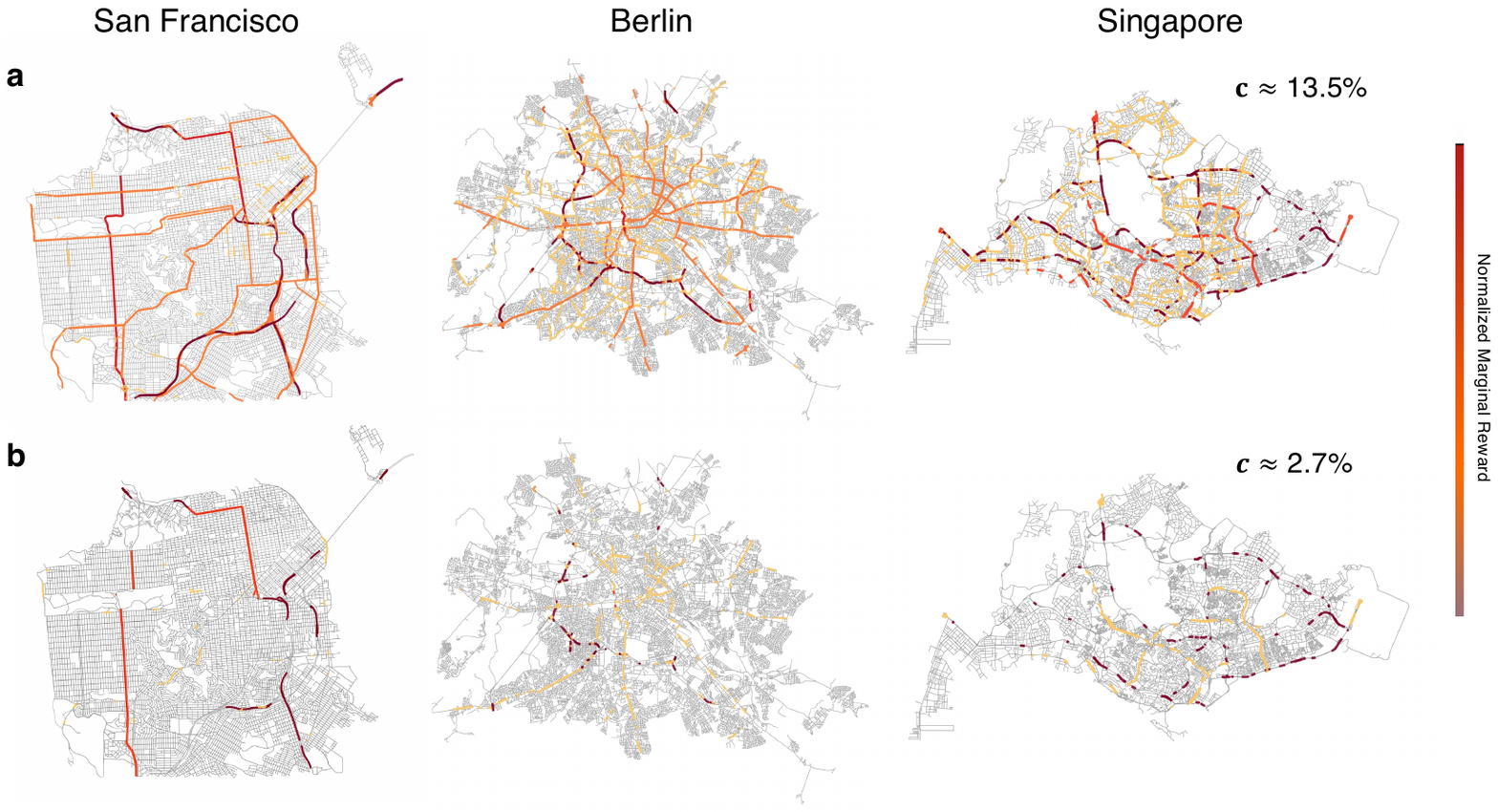}
    \caption{\textbf{Traffic-optimal DAS placement for San Francisco, Berlin, and Singapore}. \textbf{a} Coverage ratio of $\boldsymbol{c}\approx13.5\%$, translated to a fiber length budget of 250 km, 816 km, and 611 km for the three cities, respectively. \textbf{b} Coverage ratio of $\boldsymbol{c}\approx 2.7\%$, for a fiber length budget of 50 km, 163 km, and 122 km for the three cities, respectively. The streets color represents each street's marginal contribution to the optimization reward.}
    \label{fig:TrafficMonitoring}
\end{figure}

\subsection{Optimization of DDAS networks}
For practical purposes, the number of available DAS interrogators imposes a stronger limitation than fiber coverage. Assuming short-range DAS interrogators that can be placed anywhere along the network, we present optimization approaches for three use-cases: traffic flow, earthquake early warning, and activity monitoring. These optimizations are tailored to specific cities worldwide, based only on street-map data and population density. They account for a \textit{budget}~$B$ that limits the total DAS deployment, assuming each interrogator covers a single street up to a maximal sensing range of $500\,\mathrm{m}$, which is more restrictive than typical real-world scenarios.

For traffic flow optimization, we assume that DAS can only record traffic on the street it is installed in. The optimization in Eq.~\eqref{eq:opt1} reduces to a knapsack problem in which each edge has value proportional to its traffic volume $C_{ij}$ and costs $\lceil l_{ij}/t\rceil$ interrogators, where $l_{ij}$ is the length of the street. A greedy algorithm ranks edges by their traffic-per-DAS efficiency and selects in decreasing order until the budget is exhausted (Figure~\ref{fig:TrafficMonitoring}). In addition to prioritizing short, busy streets, this approach favors multi-lane roads as we assume that a single DAS fiber on either side of the road can record traffic in all lanes.

\begin{figure}
    \centering
    \includegraphics[trim={0cm 3.5cm 0cm 2cm}, clip, width=1.0\textwidth]{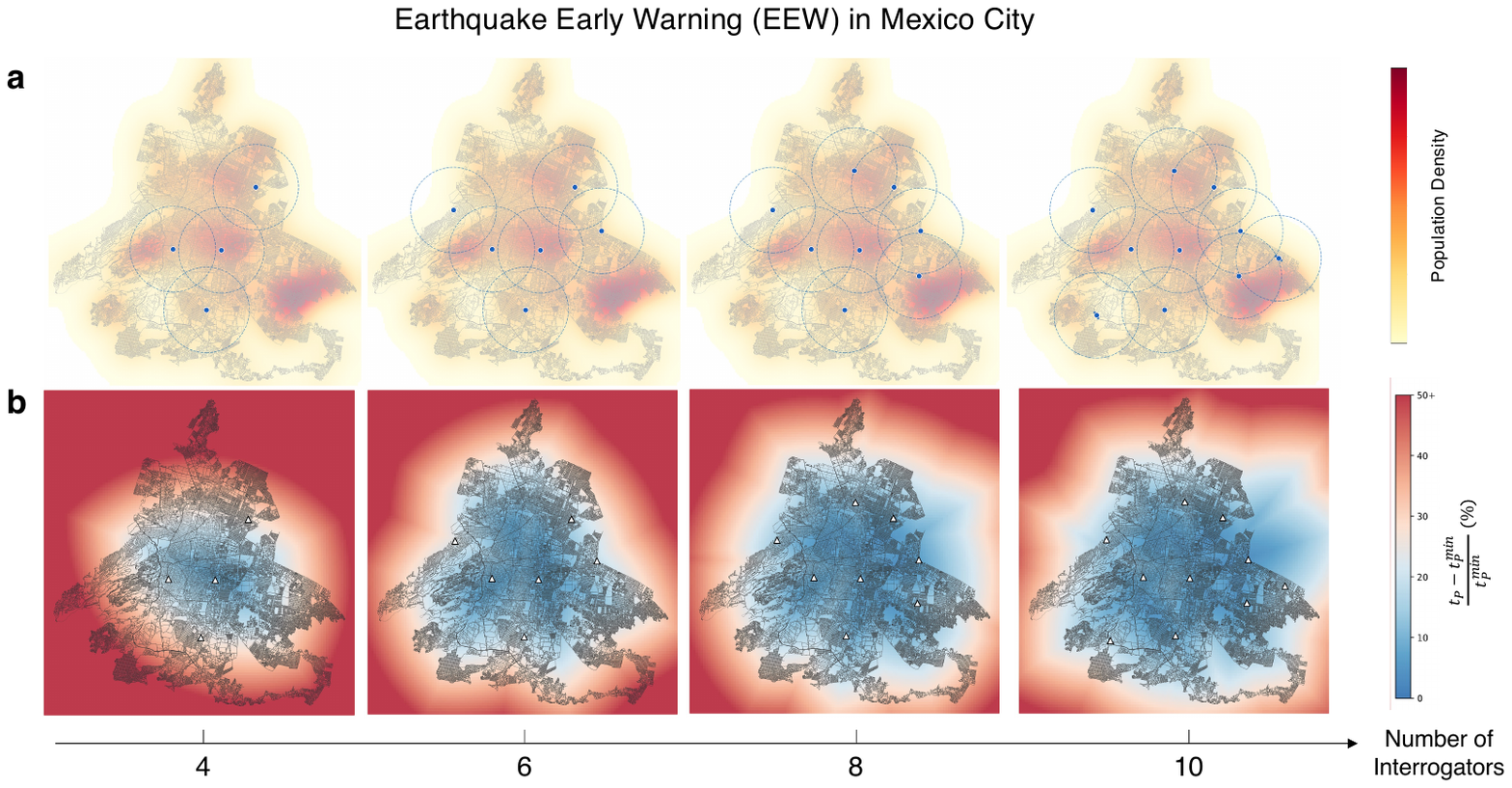}
    \caption{\textbf{Earthquake early warning in Mexico City}. The optimization is for $T = 4,\,6,\,8,\,10$ interrogators with $5\,\mathrm{km}$ minimum spacing, and $M=4$. \textbf{a} Optimal DAS placement, with the colorbar indicating population density. \textbf{b} Detection delay ratio $\eta$ (\%) at each potential epicentre (depth $h = 20\,\mathrm{km}$), indicating detection time reduction.}
    \label{fig:EWE}
\end{figure}

For earthquake early warning, we optimize DAS placement to minimize earthquake detection time. This requires P-wave arrivals at $T$ different DAS segments, which we force to be at least $D = 5\,\mathrm{km}$ apart to avoid false triggers from localized noise sources~\cite{Farghal2022early,biondi2026real}. In practice, assuming no prior knowledge of earthquake distribution, and for any potential epicentre, we minimize the P-wave arrival time at the closest $M=4$ segments (Eq.~\eqref{eq:opt2}). We linearly weigh the delay by the population density, to maximize response time in dense areas. We assume given $v_P\mathrm{ km/s}$, and an earthquake depth of $20 \: km$ for the calculations. In Figure~\ref{fig:EWE}a, we show the optimal placement of DAS interrogators in Mexico City, overlaid over population density. Figure~\ref{fig:EWE}b shows the delay, in percentage, relative to the theoretically possible earliest warning time, which occurs at the arrival of the P-wave to the surface point directly above the source. This analysis shows that a relatively limited number of synchronized DAS segments can serve as a highly effective early warning network. 

\begin{figure}
    \centering
    \includegraphics[trim={0.5cm 5.5cm 2.5cm 3cm}, clip, width=1.0\textwidth]{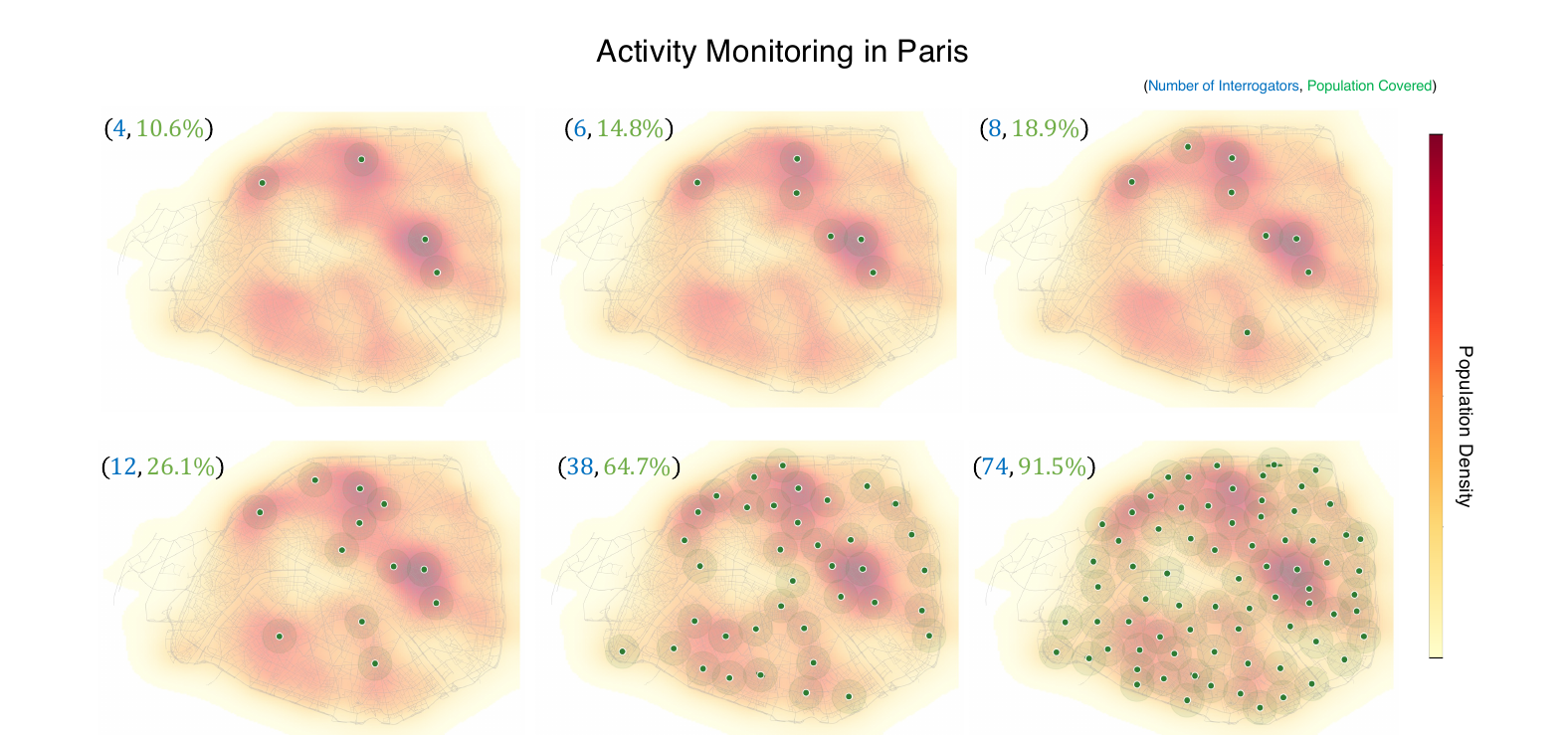}
    \caption{\textbf{Activity monitoring in Paris}. Coverage disks ($R = 600\,\mathrm{m}$) are shown for $T = 4$ to $T = 74$ interrogators, achieving $10.6\%$ to $91.5\%$ population coverage respectively.}
    \label{fig:ActivityMonitoring}
\end{figure}

For urban activity monitoring, each DAS interrogator is modeled with a $600\,\mathrm{m}$ sensing radius, and the optimization maximizes the fraction of the population within the sensing area (Eq.~\eqref{eq:opt3}). Figure~\ref{fig:ActivityMonitoring} shows results for Paris: with only $T=4$ interrogators approximately $10.6\%$ of the city's population is covered, whereas $T=74$ interrogators suffices to reach $91.5\%$ coverage. The rapid initial growth reflects the high concentration of population in the city centre, where a single well-placed sensor covers a disproportionately large share of residents.

\section{Discussion}
Currently, DAS interrogators are not fit for large-scale urban sensing due to the structure and topology of urban fiber networks. We describe a roadmap for when short-range, on-chip DAS is available and can replace traditional interrogators, thus allowing significantly more flexibility in designing an urban DDAS array that can be leveraged for different applications. We assume that every street can be interrogated by such short-range DAS. This assumption is supported by the high fiber coverage typically observed in dense urban areas. As exemplified by Tel-Aviv Jaffa, there is above $91\%$ coverage for direct customer access, and probably higher overall if we take into account governmental/municipal fibers and existing telecommunication infrastructure that does not serve non-residential areas. In addition, fiber coverage is only expected to grow with time. We thus conclude that fiber coverage will not be a limiting factor. A second assumption is that it will be possible to connect short-range DAS at every location. The exact limitations encountered in real-world deployments will depend on the design of such DAS systems, but will most likely depend on access and electrical power. In any case, the network optimization we describe here is readily adapted to a subset graph that only includes fiber-instrumented streets in which short-range DAS interrogators can be installed. 

The length of each sensing segment controls the beamforming power of the DAS array and its mitigation of noise. For subsurface imaging tasks, it also limits the depth and resolution of subsurface imaging targets. The trajectory of the fiber may also be a limiting factor for such purposes as most studies favor straight segments, which are not necessarily the case in urban environments. In our analysis, we assumed that the array length was not a limiting factor, and that perfect sensing could be achieved while recording over a single street of any length. In practice, DAS recording over several contiguous streets might be necessary to overcome low signal-to-noise conditions and better resolve subsurface structures. The exact deployment and connection strategy will depend on the capabilities of short-range DAS and the topology and components of the existing fiber network. For these reasons, certain modifications to our proposed approach will be required to guide real-world network design, but they are not expected to fundamentally change the results.

Even under the simplified scenario we study, the link between coverage ratio and achievable spatial resolution is not obvious. Our analysis shows that sensing applications typically fall into one of two regimes - low-coverage (less than $10 \%$ of the streets) or effectively complete coverage ($ \boldsymbol{c} \rightarrow 100\%$). In between, there is a phase transition in coverage, with a critical point at $\boldsymbol{c}^\star=51.6\%$. This value marks the transition from street/neighborhood-scale coverage to city-scale coverage, where the DDAS network becomes fully connected. Under this regime, traffic monitoring becomes possible, as vehicles can be detected along DAS-instrumented street as well as junctions crossing them. Pedestrian activity tracking, although more challenging, can also be conducted statistically. Crossing the percolation threshold may, in general, open the path for new applications that require aggregate sensing across an entire city. 

Our analysis shows that covering less than $10 \%$ of the streets with DAS can yield significant benefits in earthquake early warning, geological mapping, groundwater monitoring, and urban activity tracking. These approaches leverage two complementary strengths of the proposed DDAS network. First, it has a locally high (meter-scale) resolution over instrumented segments, which can be used for seismic array processing. Second, it allows an optimal distribution of sensing segments without being forced to create a contiguous optical path that requires large modifications to the fiber infrastructure. This indicates that there is immediate value in implementing DDAS networks within a timeframe of 5-10 years, depending on the rate of development of short-range DAS. 

In this study, we show particular cases that exemplify optimal network designs for a specific task. However, these are only projections of a generalized network optimization problem that takes into account the relative importance of different applications and designs the best possible network. Naturally, the end user will be required to define this weighting based on the needs of the city. In addition, our optimization approach can utilize existing a priori information about the targets of DAS sensing. For example, mapped faults, existing seismological stations, geological knowledge, hydrological measurement points and urban areas of special interest can be taken into account using appropriate weighting that indicates whether a certain area is more or less interesting for one of the applications. On the other hand, in applications for which there is no prior knowledge, the random models we analyzed will coincide with the actual network design. 

Even taking into account the current performance of DAS, which is expected to deteriorate for on-chip designs, the risk for privacy-invading applications is extremely low. Pedestrian monitoring and individual vehicle tracking require effectively full coverage with success probability which scales as $\mathcal{O}\big(\delta^2\ln(\tfrac{1}{\varepsilon})\big)$, for coverage of $1-\varepsilon$, and distance $d=\delta \ll 1$, which indicates a significant jump in the error for any coverage reduction $\varepsilon$. Achieving such high coverage is not expected in the near future. In a more distant future, designing optimal networks that still maintain privacy may become a question that will require addressing. 

The existing issues in accessing the urban fiber-optic network are expected to remain and potentially worsen. Fiber ownership is fragmented between different commercial operators as well as governmental organizations. Mapping can be highly inaccurate, especially for older installations. Concerns about data privacy, although DAS does not interfere nor can it analyze standard optical telemetry, often limit collaboration potential. For DDAS to become one, if not the backbone, of the sensing tools of the smart cities of the future, stakeholders must collectively share access, data, and information, potentially under a governmental or regulatory framework. Operating in conjunction with long-range DAS interrogators in specific segments can greatly improve performance, especially for traffic monitoring.

\section{Methods}
\subsection{Urban Street Network Model}
We represent each city's street network as a planar graph $\boldsymbol{G}(V,E)$, where nodes $V$ are street junctions and edges $E$ are street segments. This representation is constructed from OpenStreetMap data using OSMnx \cite{boeing2017osmnx}, retaining only drivable roads. The planar graph approximation is justified by the structural properties of urban street networks \cite{cardillo2006structural,barthelemy2011spatial} which exhibit an average node degree of $\langle k \rangle \approx 4$ and are well-described by Gabriel graph statistics. Specifically, a Gabriel graph is defined as a subgraph of the Delaunay triangulation where an edge $(u,v)$ exists if and only if the closed disk with diameter $uv$ contains no other nodes \cite{gabriel1969new}. This construction mimics the economic logic of urban planning — direct connections between neighbors, without redundant shortcuts — and has been shown to reproduce the statistical properties of real urban street networks\cite{barthelemy2011spatial}.

The monitored fiber subgraph $\boldsymbol{g}(V, e)$ shares the same node set, where $e \subseteq E$ represents the subset of streets instrumented with DAS. The coverage ratio $\boldsymbol{c}$ is defined as:

\begin{equation}
    \boldsymbol{c} = \frac{|e|}{|E|} \in [0, 1] \ .
\end{equation}

We assume that, to the first order, an optical fiber is present along every street, consistent with our survey of Tel Aviv-Jaffa. Under this assumption, $\boldsymbol{c}$ reduces to the fraction of streets on which a DAS interrogator is installed. We note that typically $N=|V|$, the number of nodes in the graph, is very large in urban areas, ranging from a few thousand to even millions (see Supplementary Materials). This point is important for the analysis introduced next, as percolation theory is valid under the thermodynamic limit, where $N \rightarrow \infty$. 

\subsection{Percolation and coverage analysis}
We analyze the sensing capability of $\boldsymbol{g}$ as a function of $\boldsymbol{c}$ using two complementary percolation models:

\subsubsection{City-scale connectivity}
The emergence of a Giant Connected Component (GCC) in $\boldsymbol{g}$ marks the transition from fragmented, neighborhood-scale coverage to city-wide sensing. Based on bond percolation theory on Gabriel graphs, the size of the largest cluster $S_{\text{max}}$ scales with the number of nodes $N$ according to distinct phases governed by the critical bond percolation threshold $\boldsymbol{c}^\star$ \cite{bertin2002continuum}. For Gabriel graphs, this threshold has been established at $\boldsymbol{c}^\star \approx 0.516$ \cite{norrenbrock2016percolation}. Below $\boldsymbol{c}^\star$, the largest connected sensing cluster grows only logarithmically with $N$.
Above $\boldsymbol{c}^\star$, it grows as $\left(\boldsymbol{c}-\boldsymbol{c}^\star \right )^{\frac{5}{36}} N$, enabling macroscopic city-wide coverage. The full scaling behavior, along with additional derivation details, are given in the Supplementary Materials.

\subsubsection{Sparse sensing}
For applications that do not require network-wide connectivity, and instead require a sensing fiber to exist within a distance $d$ of any target location, we derive the required coverage as follows. Assuming edges of $\boldsymbol{g}$ are distributed uniformly across $\boldsymbol{G}$, the probability of \textit{not} encountering any such edge within a given \textit{neighborhood volume} $\mathcal{V}(d)$ is given by $(1-\boldsymbol{c})^{\mathcal{V}(d)}$. The volume element $\mathcal{V}(d)$ represents the expected number of edges within $d$ hops. In planar graphs with $\langle k \rangle \approx 4$, this neighborhood volume cannot grow too fast, and it scales quadratically: $\mathcal{V}(d) \approx \alpha d^2$, with some topological expansion coefficient, $\alpha$, which dictates the local mesh density of the planar graph. $\alpha$ can be analytically approximated as half the average node degree: $\alpha \approx \langle k \rangle / 2$ \cite{barthelemy2011spatial}.

Thus, the probability $P(d)$ of having at least one monitored fiber inside the sensing neighborhood is given by $1$ minus the probability that this entire volume contains no fibers:\begin{equation}
    P(d) = 1 - (1 - \boldsymbol{c})^{2d^2} \ .
    \label{eq:coverage_prob}
\end{equation}
Inverting this relation allows us to evaluate the effective sensing resolution $d$ for a given existing coverage $\boldsymbol{c}$:
\begin{equation} \label{eq:dc}
    d(\boldsymbol{c}) = \sqrt{\tfrac{1}{2} \log_{1-\boldsymbol{c}}(1 - P_{\text{success}})} \ .
\end{equation}

where we replace $P(d)$ with $P_{\text{success}}$, the success probability of an algorithm for a given application. 

\subsection{Required Resolution for Sensing Applications}
The above formulation measures sensing distance as a function of $d$, which is a dimensionless parameter. However, we are often interested in physical distances for application in which we know the sensing range of the fiber or are interested in spatial resolution. Conversion from the dimensionless hop distance $d$ to a physical distance $r$ is given by the relation $\rho \pi r^2 = \mathcal{V}(d)$, where the density parameter $\rho$ measures the number of street edges per square kilometer. This parameter is approximated by the total number of edges divided by the total city area. In this study, $\rho \approx 99.8 \text{ km}^{-2}$, as was estimated based on 25 cities in the dataset. Therefore, $\rho \pi r^2 \approx 2d^2$ and $r=\sqrt{\frac{2}{\pi \cdot99.8}}d \approx 0.08d $, or $80 m$ as the typical distance added per unit of $d$. It should be noted that $\rho$ varies significantly between cities, with a typical standard deviation of about $55.8\,\text{km}^{-2}$. This suggests that it is preferable to estimate $\rho$ separately for each city to improve precision, although here we present the average case.. Based on existing literature, we estimate the required $d$ for different applications, converting between physical distance and graph parameter $d$. We emphasize that these are broad estimates that are meant to convey a general classification rather than strong constraints. These values are highly dependent on geological structure, DAS quality, noise characteristics, and others.  
\begin{itemize}
    \item Earthquake early warning - existing broadband or accelerometer networks have typical distances of at least several kilometers between stations, even in remote areas, hence $d=10-25$ is sufficient to outperform them \cite{Anderson2024Expansion,McGuire2025hazard}.
    \item Geological units mapping - whereas this value varies greatly, it typically changes on the scale of a kilometer. Therefore, $d\approx 7-12$ is sufficient to detect shallow large-scale structural changes \cite{Mirzanejad2025Imaging,yuan2024DAS}.
    \item Urban activity tracking - a single DAS interrogator has been shown to capture activity over a range of approximately 300-1000 m, depending on the type of activity (traffic, construction, industrial activity, school-related). Therefore, $d\approx 3-8$ is sufficient for a full city coverage \cite{liu2025urban}.
    \item Groundwater monitoring - large-scale behavior changes are on the scale of $500 \,\mathrm{m}$ and above, especially dependent on whether groundwater is stored in shallow surficial or deep aquifer and infiltration rates that depends on surface conditions. Therefore, we assume $d\approx 5-10$\cite{Li2025groundwater,Tribaldos2021aquifer,Shen2025dynamics,shen2024vadose}.
    \item Traffic monitoring - Detecting vehicles is generally possible only for the street the fiber is installed on. However, it can be conducted on both sides of the street and can detect vehicle crossings at junctions \cite{Lindsey2020COVID,cohen2025fiber}. Therefore, $d\approx 0.1-0.9$ is sufficient for a full coverage of the city, because even if a street is not covered, vehicles can potentially be detected at its intersections.
    \item Individual vehicle tracking - To track an individual vehicle, fiber along the entire route is required and hence $d \approx 0-0.05$.
    \item Infrastructure monitoring - possible only if the fiber is directly installed on the monitoring target or very close to it, so $d\approx 0-0.1$ \cite{Ghazali2024Civil,rodet2025urban,liu2023bridge}.
    \item Pedestrian monitoring - has a significantly lower sensing range than for vehicles, thus $d\approx 0-0.05$ \cite{luckie2025footsteps}.
\end{itemize}

\subsection{DDAS Network Optimization}
In this section, we describe the optimal design of a graph $g$ for a given urban area, for which we know the street map and population density, under a constraint on the number of interrogators. We present three particular cases: traffic monitoring, earthquake early warning, and urban activity tracking.
\subsubsection{Traffic Monitoring}
For the case of traffic management, the optimal routing monitors as many vehicles per unit time as possible. We define $C_{ij}$ as the traffic flow (vehicles per unit time) from junction~$i$ to junction~$j$. We assume its a priori knowledge from road type mapping (as we did in this study, see Supplementary Materials) or mobility/camera data which would be a better choice if possible. The cost of DAS interrogation can be defined in different ways that ultimately depend on the design of short-range DAS sensing and available fiber-optic network. In this study, we assume that each DAS interrogator has a finite sensing range $t = 500 \,\mathrm{m}$; a street of length~$l_{ij}$ therefore requires $\lceil l_{ij}/t \rceil$ interrogators to be monitored. The cost function we use is the total number of interrogators. Nonetheless, other cost metrics such as total monitored length, number of needed access points, or any combination of those can be used with the same framework. The budget $B$ imposes an upper bound $\mathrm{Price}(\boldsymbol{g})\leq B$.

\begin{gather} \label{eq:opt1}
    \mathcal{L} = \max_{\boldsymbol{g}} \sum_{e(ij)\in \boldsymbol{g}} C_{ij}
    \quad \text{s.t.} \quad \mathrm{Price}(\boldsymbol{g}) \leq B \, .
\end{gather}

This problem is a variant of the 0-1 knapsack problem: each edge has value $C_{ij}$ and its cost, the number of interrogators. We solve it with a greedy algorithm that ranks edges by their efficiency $\eta_{ij} = \tfrac{C_{ij}}{\lceil l_{ij}/t\rceil}$ and selects in decreasing order until the budget is exhausted. This naturally favors short, high-traffic streets. The ability of DAS to detect vehicles on multiple lanes and over both driving directions pulls the solution towards wider roads. 

\subsubsection{Earthquake Early Warning (EEW)}

For earthquake early warning, the objective is to position DAS interrogators such that any earthquake source can be rapidly detected by multiple, spatially separated sensing segments. This is important to avoid false alarms, provide an initial location of the earthquake, and reliably estimate its magnitude \cite{McGuire2025hazard}. We assume an earthquake with surface epicentre $\mathbf{q}$ and hypocentral depth $h$. The P-wave propagates at velocity $v_P$ and reaches sensor located at $s_i$ after traveling the three-dimensional hypocentral distance
\begin{equation}
    \delta_i(\mathbf{q}) = \sqrt{\lVert \mathbf{s}_i - \mathbf{q}\rVert^2 + h^2}\,.
\end{equation}
Assuming a constant P-wave velocity. Reliable detection requires the P-wave to reach at least $M$ segments, among $T$, the total number of segments, and the detection time is therefore governed by $\delta_{(M)}(\mathbf{q})$, the $M$-smallest hypocentral distance among all selected sensors. We assume $M=4$, a fixed depth of $h=20 \: km$, and enforce a minimal distance $D=5 \: km$ between any two segments. These are typical values that can be changed based on the local seismic hazard assessment and known noise sources. The optimization then aims to minimize the earthquake detection time:

\begin{gather} \label{eq:opt2}
    \mathcal{L} = \min_{\mathbf{s}} \; \mathbb{E}_{\mathbf{q}\sim \boldsymbol{\rho}(\mathbf{q})} \; \bigg[ \delta_{(M)}(\mathbf{q},\,\mathbf{s}) \bigg]
    \quad \text{s.t.} \quad
    \lVert s_i - s_j \rVert \geq D \;\;\forall\, i \neq j,
    \quad \mathrm{Price}(\boldsymbol{g}) \leq B \,.
\end{gather}

We quantify detection quality by the \emph{detection delay ratio}, $\eta(\mathbf{q}) = \frac{\delta_{(M)}(\mathbf{q})}{h}$. It equals $100\%$ at the time that the earthquake first reaches the surface, which is the theoretically possible earliest detection time ignoring signal-to-noise ratio, ambiguity, and the need for location and magnitude estimation. The optimal sensor placement minimizes the worst-case Mth-nearest-sensor hypocentral distance over all possible epicentres. Since the epicentre location is unknown a priori, possible earthquake locations are sampled from a probability density function $\boldsymbol{\rho}(\mathbf{q})$, which matches the normalized population density, thereby prioritizing coverage in densely populated areas. 

\subsubsection{Activity Monitoring}
In urban activity monitoring, each DAS segment is modeled as having a certain range, measured by Euclidean surface distance, in which it can detect activity. We model it as a radius of coverage $R$ around the midpoint of each selected edge. In practice, the type of activity, fiber coupling, and noise levels control $R$, but we assume it is constant over the entire graph $\boldsymbol{g}$. The objective is to select edges that collectively cover as large a fraction of the urban population (known from WorldPop density data \cite{worldpop2018}) as possible:
\begin{gather} \label{eq:opt3}
    \mathcal{L} = \max_{\mathbf{s}} \; \int_{\mathcal{B}} \rho(\mathbf{x}) \, \mathrm{d}\mathbf{x}
    \quad \text{s.t.} \quad
    \mathrm{Price}(\boldsymbol{g}) \leq B \,,
\end{gather}

where $\mathcal{B} \equiv \bigcup_i \mathcal{B}(s_i,\, R)$, and $\mathcal{B}(s_i, R) = \{ \mathbf{x} \in \mathbb{R}^2 : \lVert \mathbf{x} - s_i \rVert \leq R \}$ is the coverage disk of sensor $s_i$, and $\boldsymbol{\rho}$ is the population density. This is a variant of the maximum coverage problem. We precompute a coverage matrix whose entry~$(i,j)$ is unity whenever $\lVert s_i - s_j\rVert \leq R$, then apply a greedy algorithm that iteratively selects the edge whose coverage disk captures the largest uncovered population. The algorithm could be extended to include a range-dependent sensing probability as well as handling different types of activities with different sensing ranges.

\backmatter

\section*{Declarations}

\textbf{Funding} 
K.C. gratefully acknowledges the Milner Foundation, and the VATAT (PBC) Fellowship for Outstanding PhD Students in Data Science. This research was supported by the Israeli Ministry of Science, Technology, and Space under grant no 1001953425, the Blavatnik Artificial Intelligence and Data Science Fund, and the Shlomo Shmeltzer Institute for Smart Transportation at Tel Aviv University.

\noindent \textbf{Conflict of interest}
The authors declare no competing interests. 

\noindent \textbf{Ethics approval}
Not applicable. This study utilizes pre-existing infrastructure and does not involve direct interaction with human subjects.

\noindent \textbf{Consent for publication}
Not applicable.

\noindent \textbf{Data availability}
All underlying data used in this study are publicly available and are cited from OpenStreetMap \cite{openstreetmap}, WorldPop \cite{worldpop2018}, and Urban Street Networks \cite{boeing2020indicators}.

\noindent \textbf{Code availability}
The custom Python code implementing the optimizations, together with the urban street-network and population datasets used in this
study, is available in the GitHub repository~\cite{cohen2026das}.

\noindent \textbf{Author contribution}
K.C. and A.L. contributed equally to all aspects of this work.

\bibliography{sn-bibliography}

\newpage
\appendix











\section{City size statistics}
\label{app:citySizeStat}
\begin{figure}[b]
  \centering
  \includegraphics[trim={0cm 5cm 0cm 5cm}, clip, width=1.0\textwidth]{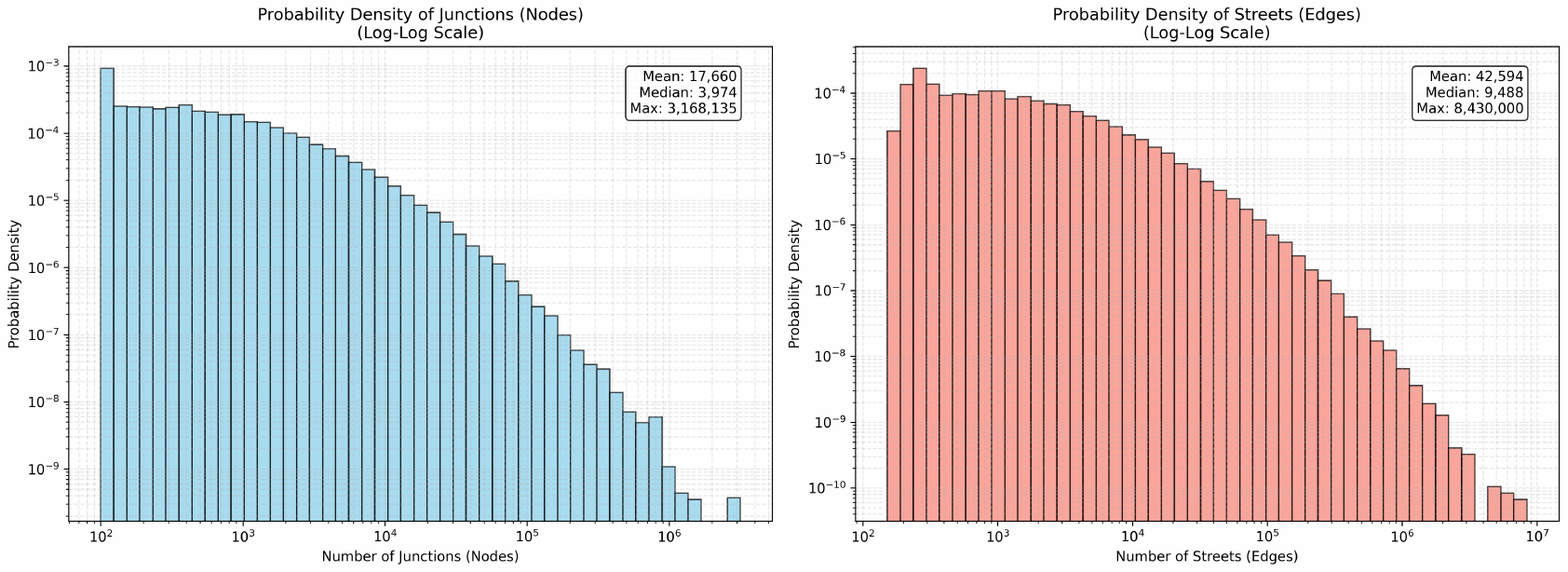}
  \caption{\textbf{Probability density of urban street network sizes (log--log scale) for over 8{,}900 cities worldwide}. Left: number of junctions (nodes). Right: number of street segments (edges). Both distributions are heavy-tailed, indicating that most cities have   small networks while a few mega-cities contain orders of magnitude more streets.}
  \label{fig:city_distributions}
\end{figure}

Figure \ref{fig:city_distributions} shows the distribution of street network sizes across over $8,900$ urban areas worldwide, based on the Boeing global street network dataset~\cite{boeing2017osmnx}. Both the number of junctions (nodes) and street segments (edges) follow a heavy-tailed distribution spanning several orders of magnitude, from small towns with fewer than ten junctions to mega-cities with millions of street segments. The average city has approximately 17{,}660 junctions, while the largest networks exceed $10^6$.

For the traffic-monitoring optimisation, each street segment is assigned a surrogate traffic weight based on its OpenStreetMap \texttt{highway} tag (Table~\ref{tab:highway_weights}).  The weights follow a roughly geometric progression that mirrors the functional road hierarchy: higher-class roads (motorways, trunks) serve regional through-traffic and carry substantially higher volumes, whereas residential and service streets primarily provide local access.  Because the greedy knapsack ranking depends only on the \emph{relative} ordering of edge efficiencies, the optimization result is invariant to any positive rescaling of the weight vector; accordingly, the specific magnitudes serve as ordinal proxies rather than calibrated vehicle counts.

\begin{table}
  \centering
  \caption{Surrogate traffic weights assigned to street segments by
    OpenStreetMap \texttt{highway} classification.  Values decrease
    roughly geometrically with functional road class, reflecting the
    well-documented correlation between road hierarchy and traffic
    volume. Link types inherit the weight of the parent class (or a
    fraction thereof).}
  \label{tab:highway_weights}
  \begin{tabular}{llr}
    \toprule
    \textbf{Highway tag} & \textbf{Road function} & \textbf{Weight} \\
    \midrule
    \texttt{motorway}        & Controlled-access highway     & 2000 \\
    \texttt{trunk}           & Major national/regional road   & 1500 \\
    \texttt{primary}         & Primary urban arterial         & 1000 \\
    \texttt{secondary}       & Secondary arterial             &  500 \\
    \texttt{tertiary}        & Tertiary / collector street    &  250 \\
    \texttt{unclassified}    & Minor through-road             &  100 \\
    \texttt{residential}     & Local residential access       &   50 \\
    \texttt{living\_street}  & Shared-space / traffic-calmed  &   20 \\
    \texttt{service}         & Parking, alleys, driveways     &   10 \\
    \midrule
    \texttt{motorway\_link}  & Motorway on/off ramp           & 1000 \\
    \texttt{trunk\_link}     & Trunk on/off ramp              &  750 \\
    \texttt{primary\_link}   & Primary intersection link      &  500 \\
    \texttt{secondary\_link} & Secondary intersection link    &  300 \\
    \texttt{tertiary\_link}  & Tertiary intersection link     &  150 \\
    \texttt{road}            & Unknown classification         &   50 \\
    \bottomrule
  \end{tabular}
\end{table}

\section{Tel Aviv Municipality Fibers Coverage}
\label{app:TAcoverageExp}

To validate that urban fiber-optic coverage is sufficiently dense to support our analysis, we conducted a survey of 433 randomly sampled addresses in Tel Aviv, Israel (Table~\ref{tab:fiber_validation}). For each address, fiber availability was determined via the Israeli telecommunication companies' broadband coverage portal. Addresses were geocoded using the OpenStreetMap Nominatim service and filtered to retain only those falling within the Tel Aviv municipal boundary.
Of the 410 addresses successfully geolocated within the city, 373 (91.0\%) were found to have at least one fiber-optic provider, confirming that fiber infrastructure in Tel Aviv is sufficiently pervasive for the percolation-based analysis presented in this work.

\begin{table}[t]
  \centering
  \caption{Summary of the fiber-optic availability survey   across 433 randomly sampled addresses in Tel Aviv, Israel.   Fiber availability was determined by querying each address on   the Israeli telecommunication companies' broadband coverage portal.}
  \label{tab:fiber_validation}
  \begin{tabular}{l r r}
    \toprule
    & Count & Percentage \\
    \midrule
    Addresses surveyed              & 433 &        \\
    Successfully geocoded           & 424 &        \\
    Within Tel Aviv municipality    & 410 &        \\
    \midrule
    Fiber available                 & 373 & 91.0\% \\
    No fiber                        &  37 &  9.0\% \\
    \bottomrule
  \end{tabular}
\end{table}

\section{Additional Mathematical derivations}
\label{app:math}

\subsubsection*{The Percolation Model}
As $\boldsymbol{c}$ describes the average coverage along the city's streets, it is clear that too small a value is not sufficient to provide large-scale monitoring. Therefore, we expect a universal threshold that models the transition from localized to city-wide monitoring, and this threshold corresponds to the emergence of a Giant Connected Component (GCC) in the fiber subgraph.

Based on bond percolation theory on planar graphs, the size of the largest connected cluster, $S_{\text{max}}$, scales with the total number of network nodes $N$ according to distinct structural phases. These phases are governed by the critical percolation threshold $c^\star$ \cite{bertin2002continuum}. For Euclidean Gabriel graphs, which belong to the standard 2D percolation universality class, this threshold is numerically established at $\boldsymbol{c}^\star \approx 0.5167$ \cite{norrenbrock2016percolation}. 

The scaling behavior of the largest cluster is dictated by the critical exponents $\beta = 5/36$ and $\nu = 4/3$, yielding:
\begin{equation} \label{eq:GCG_Smax}
    S_{\text{max}}(\boldsymbol{c}) \sim 
    \begin{cases}
        \ln N & \boldsymbol{c} < \boldsymbol{c}^\star \quad \text{(subcritical)} \\
        N^{91/96} & \boldsymbol{c} \approx \boldsymbol{c}^\star \quad \text{(critical)} \\
        (\boldsymbol{c} - \boldsymbol{c}^\star)^{5/36} N & \boldsymbol{c} > \boldsymbol{c}^\star \quad \text{(supercritical)}
    \end{cases} \ .
\end{equation}

The critical transition is further characterized by the divergence of the correlation length $\xi$ and the order parameter fluctuations $\chi$. Measuring $\xi$ strictly via topological distance (edge hops) accurately captures the true structural connectivity of the network as it approaches criticality. As the bond coverage approaches $\boldsymbol{c}^\star$, these quantities scale according to their respective universal exponents:
\begin{equation} \label{eq:GCG_divergence}
    \xi(\boldsymbol{c}) \sim |\boldsymbol{c} - \boldsymbol{c}^\star|^{-\nu}, \quad \chi(\boldsymbol{c}) \sim |\boldsymbol{c} - \boldsymbol{c}^\star|^{-\gamma} \ .
\end{equation}
where $\gamma = 43/18$. Numerical finite-size scaling analyses on Gabriel graphs verify this universal behavior, yielding $\nu \approx 1.33$ and $\gamma \approx 2.39$ \cite{norrenbrock2016percolation}. At the critical point $\boldsymbol{c} \approx \boldsymbol{c}^\star$, the structural volatility $\chi$ of the finite clusters reaches a sharp maximum, mathematically marking the exact boundary where localized, finite components rapidly amalgamate into the macroscopic spanning graph.

\begin{figure}[t]
    \centering
    \includegraphics[width=0.7\linewidth]{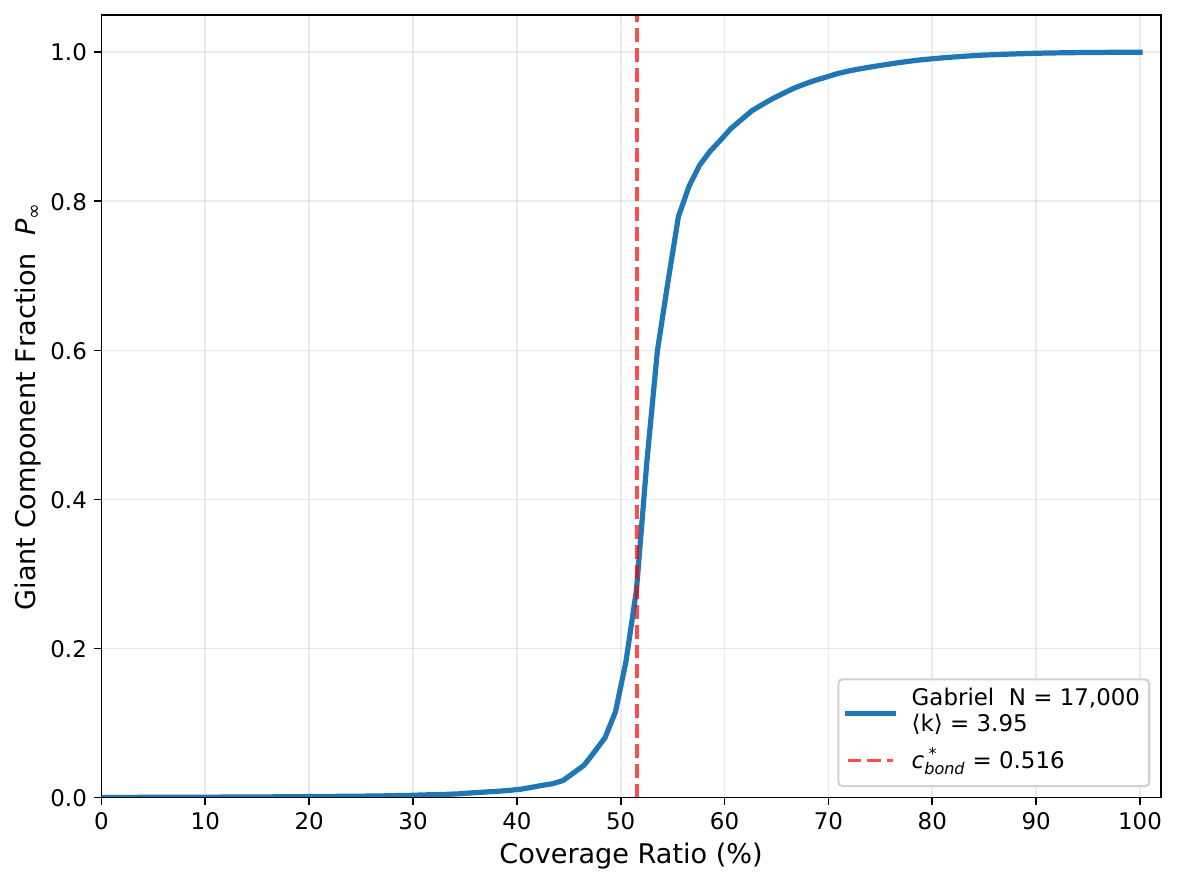}
    \caption{\textbf{Bond percolation on a Gabriel-graph model of an urban street network.} A Gabriel graph is generated on $N = 17{,}000$ points, yielding a mean degree $\langle k \rangle = 3.95$. Each edge is independently retained with probability $\boldsymbol{c}$ (the fiber-coverage ratio), and the giant connected component fraction $P_\infty$ is plotted against $\boldsymbol{c}$, averaged over $15$ independent percolation realizations. The red dashed line marks the theoretical bond-percolation threshold $c^*_{\mathrm{bond}} = 0.516$.}
    \label{fig:percolation_gabriel}
\end{figure}

Figure \ref{fig:percolation_gabriel} shows a percolation example for the Gabriel Graph, the model we chose to represent urban cities. The figure clearly illustrates how the system exhibits a sharp, continuous phase transition: below the threshold, the network fragments into disconnected clusters, while above it, a system-spanning component emerges.

\subsubsection*{Individual Vehicle Tracking}
While tracking a single vehicle along an optical fiber has already been demonstrated and shown to work well, continuous tracking at a city scale is a much more complicated task, since it requires switching between different optical fibers. Here, we assume that a tracking algorithm along a single optical fiber exists and works well, and that the tracking algorithm can switch between interrogators to track vehicles along different streets.

For this model, we analyze the connectivity along the shortest path $S_{AB}$ between two random nodes $A$ and $B$. A tracking failure occurs when the vehicle traverses an edge that is not in the subgraph $g$. We define the number of tracking gaps (jumps) $J$ as
\begin{equation}
    J = \big| \{ s \mid s \in S_{AB} \text{ and } s \notin \boldsymbol{g} \} \big|.
\end{equation}
Intuitively, for a single path on the graph between $A$ and $B$, $J$ counts how many streets are not covered by the DAS-based tracker.

For a path of length $|S_{AB}|$ (the number of edges between $A$ and $B$), the probability of encountering exactly $J$ gaps follows a binomial distribution $B(|S_{AB}|, 1-\boldsymbol{c})$. The probability of successful tracking with at most $K$ allowed gaps is
\begin{equation} \label{eq:trackProb}
    P(J \leq K) = \sum_{i=0}^{K} \binom{|S_{AB}|}{i} (1-\boldsymbol{c})^i \boldsymbol{c}^{|S_{AB}| - i}.
\end{equation}
For sufficiently large graphs, the shortest path length in a Gabriel graph scales as $|S_{AB}| \approx \tau \sqrt{N}$, where $\tau$ is a normalization factor. The expected number of gaps is clearly $\mathbb{E}_{A,B}[J] = |S_{AB}|(1-\boldsymbol{c})$, where the expectation is taken over all pairs of random points $A$ and $B$.

We introduce a tolerance threshold $K$, which intuitively represents, for a single trajectory between point $A$ and point $B$, how many streets along the way we can afford to lose tracking on.
In the large-graph limit, a sharp phase transition for tracking occurs when the expected number of gaps exceeds the tolerance threshold $K$. Substituting the scaling relation, we obtain the critical coverage $\boldsymbol{c}_{\text{track}}$ required for near-continuous tracking:
\begin{equation}
    \boldsymbol{c}_{\text{track}} \approx 1 - \frac{K}{\tau \sqrt{N}}.
\end{equation}

\begin{figure}
    \centering
    \includegraphics[trim={2cm, 4cm, 2cm, 3cm}, clip, width=1.0\linewidth]{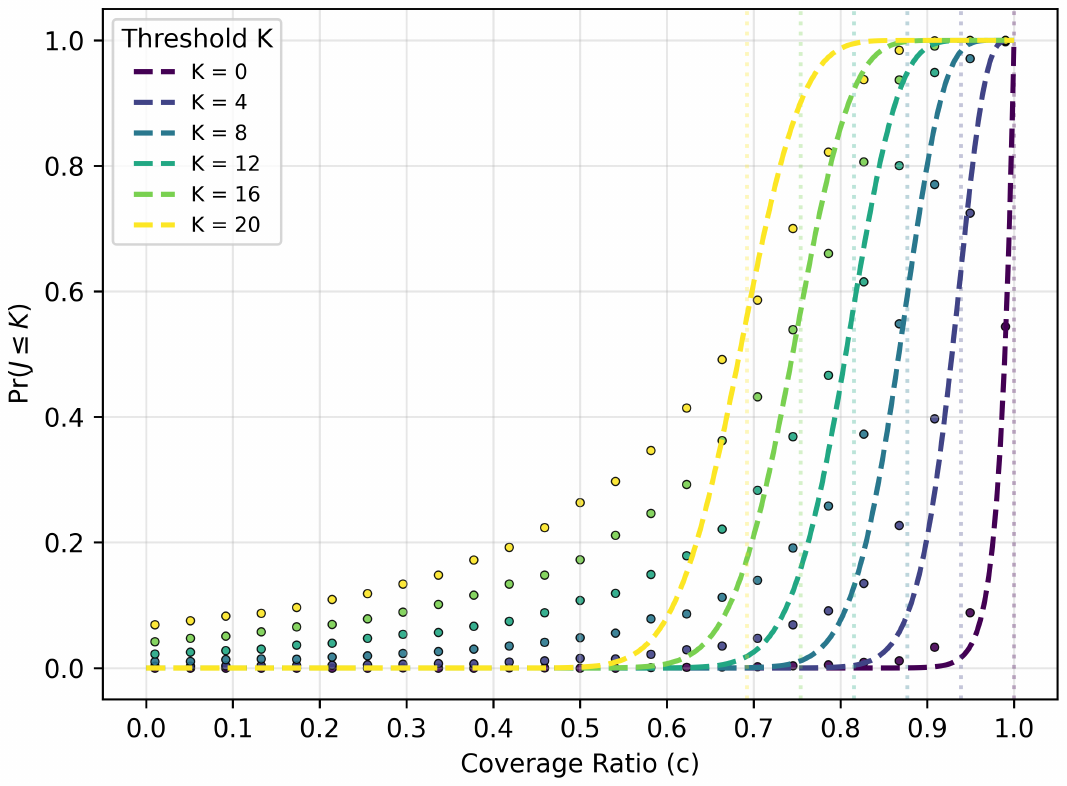}
    \caption{\textbf{Tracking probability percolation for different values of $K$}. The dots show simulation results on a random Gabriel graph with $N = 17{,}000$ nodes, while the dashed lines represent the theoretical relation predicted by Eq. \ref{eq:trackProb}.}
    \label{fig:individualTracking}
\end{figure}

Figure \ref{fig:individualTracking} shows a few simulation results for different $K$.

\end{document}